\documentclass[twocolumn,aps,prl,superscriptaddress,longbibliography]{revtex4-2}
\usepackage[colorlinks=true, citecolor=blue, urlcolor=blue, linkcolor=red]{hyperref}
\renewcommand{\section}[1]{{\par\it #1.~~}\ignorespaces}
\usepackage{amsmath,amssymb,tikz,scalerel}
\usetikzlibrary{svg.path}
\definecolor{orcidlogocol}{HTML}{A6CE39}
\tikzset{orcidlogo/.pic={
		\fill[orcidlogocol] svg{M256,128c0,70.7-57.3,128-128,128C57.3,256,0,198.7,0,128C0,57.3,57.3,0,128,0C198.7,0,256,57.3,256,128z};
		\fill[white] svg{M86.3,186.2H70.9V79.1h15.4v48.4V186.2z}
		svg{M108.9,79.1h41.6c39.6,0,57,28.3,57,53.6c0,27.5-21.5,53.6-56.8,53.6h-41.8V79.1z M124.3,172.4h24.5c34.9,0,42.9-26.5,42.9-39.7c0-21.5-13.7-39.7-43.7-39.7h-23.7V172.4z}
		svg{M88.7,56.8c0,5.5-4.5,10.1-10.1,10.1c-5.6,0-10.1-4.6-10.1-10.1c0-5.6,4.5-10.1,10.1-10.1C84.2,46.7,88.7,51.3,88.7,56.8z};}}
\newcommand\orcid[1]{\href{https://orcid.org/#1}{\mbox{\scalerel*{\begin{tikzpicture}[yscale=-1,transform shape]\pic{orcidlogo};\end{tikzpicture}}{|}}}}

\begin{document}
\title{Floquet composite Dirac semimetals}
\author{Hong Wu\orcid{0000-0003-3276-7823}}\email{Contact author: wuh@cqupt.edu.cn}
\affiliation{School of Electronic Science and Engineering, Chongqing University of Posts and Telecommunications, Chongqing 400065, China}
\affiliation{Chongqing Key Laboratory of Dedicated Quantum Computing and Quantum Artificial Intelligence, Chongqing 400065, China}
\affiliation{Institute for Advanced Sciences, Chongqing University of Posts and Telecommunications, Chongqing 400065, China}
\author{Jia-Ji Zhu}
\affiliation{School of Electronic Science and Engineering, Chongqing University of Posts and Telecommunications, Chongqing 400065, China}
\affiliation{Chongqing Key Laboratory of Dedicated Quantum Computing and Quantum Artificial Intelligence, Chongqing 400065, China}
\affiliation{Institute for Advanced Sciences, Chongqing University of Posts and Telecommunications, Chongqing 400065, China}
\author{Jian Li}
\affiliation{School of Electronic Science and Engineering, Chongqing University of Posts and Telecommunications, Chongqing 400065, China}
\affiliation{Chongqing Key Laboratory of Dedicated Quantum Computing and Quantum Artificial Intelligence, Chongqing 400065, China}
\affiliation{Institute for Advanced Sciences, Chongqing University of Posts and Telecommunications, Chongqing 400065, China}
\author{Xue-Min Yang}
\affiliation{School of Electronic Science and Engineering, Chongqing University of Posts and Telecommunications, Chongqing 400065, China}
\affiliation{Chongqing Key Laboratory of Dedicated Quantum Computing and Quantum Artificial Intelligence, Chongqing 400065, China}
\affiliation{Institute for Advanced Sciences, Chongqing University of Posts and Telecommunications, Chongqing 400065, China}
\author{Jiang-Shan Chen}
\affiliation{School of Electronic Science and Engineering, Chongqing University of Posts and Telecommunications, Chongqing 400065, China}
\affiliation{Chongqing Key Laboratory of Dedicated Quantum Computing and Quantum Artificial Intelligence, Chongqing 400065, China}
\affiliation{Institute for Advanced Sciences, Chongqing University of Posts and Telecommunications, Chongqing 400065, China}
\author{Mu Zhou}\email{Contact author: zhoumu@cqupt.edu.cn}
\affiliation{School of Electronic Science and Engineering, Chongqing University of Posts and Telecommunications, Chongqing 400065, China}
\affiliation{Chongqing Key Laboratory of Dedicated Quantum Computing and Quantum Artificial Intelligence, Chongqing 400065, China}
\affiliation{Institute for Advanced Sciences, Chongqing University of Posts and Telecommunications, Chongqing 400065, China}
\begin{abstract}

Dirac semimetals can be classified into types I, II, and III based on the topological charge of their Dirac points. If a three-dimensional (3D) system can be sliced into a family of $k_z$-dependent normal and topological insulators, type I Dirac points separate a 2D normal insulator from a 2D first-order topological insulator, while type II (III) Dirac points separate a 2D normal (first-order) insulator from a 2D second-order topological insulator. To investigate the effects arising from the interplay of distinct Dirac points, one may wonder whether these Dirac points can coexist in a single system. Here, we propose a scheme to induce composite Dirac semimetals by a special Floquet driving that preserves time-reversal and space-inversion symmetries. A general description is established to characterize Dirac semimetals in Floquet systems. The results show that Dirac semimetals hosting coexisting type I, II, and III Dirac points can be induced by delta-function or harmonic driving. Our results provide a promising new avenue for exploring novel Dirac semimetals.

\end{abstract}
\maketitle
\section{Introduction}
Owing to their potential applications in electronic devices and the new paradigm they bring to condensed matter physics, topological insulators have garnered widespread attention \cite{RevModPhys.83.1057,RevModPhys.91.015005}. The core feature of
these phases is a full insulating gap in the bulk and stable edge states protected by various internal and space group symmetries \cite{RevModPhys.88.035005,PhysRevResearch.3.013052}. According to the principle of bulk-boundary correspondence, the number of boundary states can be dictated by the corresponding topological invariants \cite{RevModPhys.88.035005}. 

Topological phases also occur in gapless systems \cite{RevModPhys.90.015001,RevModPhys.93.025002,PhysRevB.111.155115,Jiang_2021}. Gapless systems featuring symmetry-protected band crossings are termed topological semimetals. They can be divided into nodal point \cite{PhysRevLett.108.266802,PhysRevLett.125.146401,PhysRevLett.125.266804,PhysRevLett.108.140405,Bouhon_2020}, line \cite{PhysRevB.84.235126,PhysRevLett.118.045701,PhysRevLett.117.087402,PhysRevLett.132.066601,1qkj-63m8,PhysRevResearch.7.023195}, and surface \cite{PhysRevB.97.115125,PhysRevLett.132.186601,PhysRevB.111.045302} semimetals. According to topological band theory, the topological charge of band crossings can be defined over a sphere enclosing each nodal point/line/surface \cite{PhysRevLett.125.126403}. Topological edge states can be observed under open boundary conditions applied along an appropriate direction. All edge states at the Fermi level contribute to Fermi arcs connecting nodal points, lines, and surfaces with opposite topological charges \cite{PhysRevLett.125.126403,PhysRevMaterials.7.L051202,PhysRevB.110.L121118,PhysRevB.108.195306}. 

\begin{figure}[tbp]
\centering
\includegraphics[width=1\columnwidth]{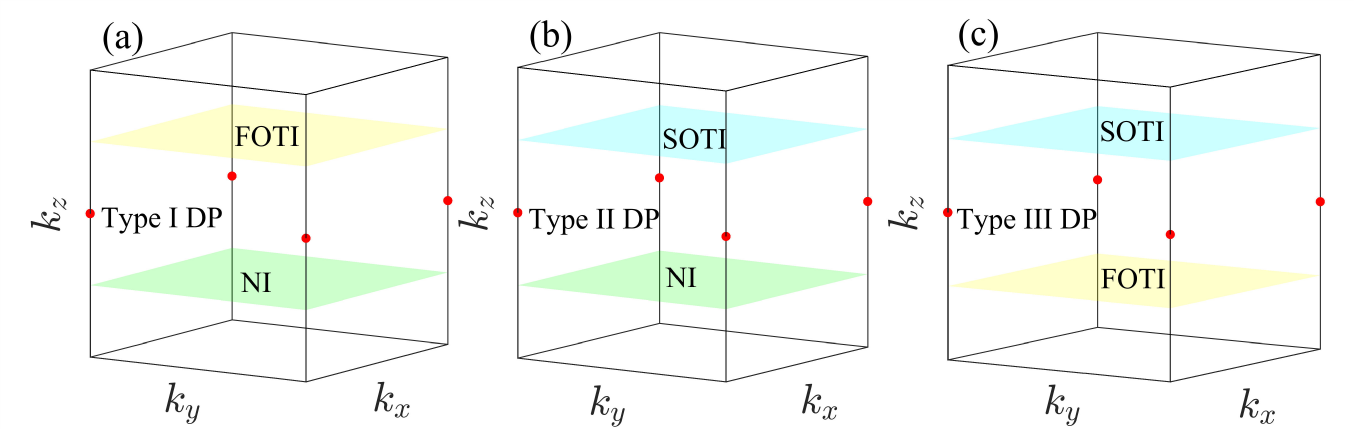}
\caption{ The Schematic of of (a) type I Dirac, (b) type II Dirac, and (c) type III Dirac semimetals. DP denotes Dirac point. Green, yellow, and blue surfaces denote trivial (NI), first-order topological (FOTI), and second-order topological (SOTI) insulators in 2D $k_z$-dependent subsystems, respectively. 
} \label{traj1}
\end{figure}

Among these gapless phases, Dirac semimetals have been widely studied in systems with crystal symmetries \cite{PhysRevLett.113.027603,PhysRevLett.127.146601,PhysRevLett.128.066602,PhysRevLett.132.219601,PhysRevB.111.195429,PhysRevResearch.7.013280,PhysRevB.111.064101,PhysRevB.111.035143,PhysRevB.111.L201408}. It features linear energy momentum dispersion around the band crossing point, called Dirac point \cite{PhysRevB.90.115207}. Due to the very high mobility \cite{Neupane_2014}, giant magnetoresistance \cite{PhysRevB.105.075141}, diamagnetism \cite{PhysRevLett.128.027201},  and low dissipation Dirac transport \cite{PhysRevB.88.125427}, this type of state can be used to design new devices, such as optical modulators \cite{Sun_2016}, ultrafast laser \cite{PhysRevB.111.064101}, and so on. All these stem from the non-trivial topology associated with Dirac points. \cite{PhysRevB.111.064101}. To clarify the topology of Dirac points, they can be classified as type I, II, and III (see Fig. \ref{traj1}). If a 3D system can be sliced into a family of $k_z$-dependent normal and topological insulators, then type I Dirac points  separate a 2D normal insulator from a 2D first-order topological insulator \cite{PhysRevB.89.235127}, while type II (III) Dirac points separate a 2D normal (first-order) insulator from a 2D second-order topological insulator \cite{Wei_2021,xz4g-jlbg}. The topological charge of any nodal structure (point, line, or surface) is given by the difference in topological invariants between the phases it separates \cite{PhysRevLett.125.126403}. Due to the different topology in type I, II, and III Dirac points, they are studied in different systems. To investigate the effects arising from the interplay of distinct Dirac points, one may wonder whether these Dirac points can coexist in a single system.  However, composite Dirac semimetals with coexistence of type I, II, and III Dirac points have not been reported. Such composite phases require spatially extended tunneling. Although this kind of hopping can be easily implemented in classic electrical circuits \cite{Song_2022}, it's difficult to realize in quantum systems

In this work, we propose a scheme for artificially creating 3D composite Dirac semimetals that host coexisting type-I, II, and III Dirac points via Floquet engineering. A general description is established to characterize Dirac semimetals in Floquet systems. Both the location of
Dirac points and the topological invariant can be determined within the theoretical framework. Based on this, we find that composite Dirac semimetals can be induced by delta-function driving in the time domain or harmonic driving. Therefore, our proposal suggests a degree of universality. Our results enrich the family of topological states and can also provide a platform to research potential applications.

\section{Floquet semimetals}
For a time-periodic system $\mathcal{H}(t)=\mathcal{H}(t+T)$, it does not have a well-defined energy spectrum. With the help of Floquet theorem \cite{PhysRevLett.111.175301}, the one-period evolution operator ${U}(T)=\mathbb{T}e^{-i\int_{0}^{T}\mathcal{H}(t)dt}$ defines an effective Hamiltonian $\mathcal{H}_\text{eff}\equiv {i\over T}\ln [{U}(T)]$ whose eigenvalues are called quasienergies. From the eigenvalue equation $U({T})\lvert u_l \rangle=e^{-i\mathcal{E}_{l}T}\lvert u_l \rangle$, we conclude that quasienergy $\mathcal{E}_l$ is a phase factor, which is defined modulus $2\pi/T$ and takes values in the first quasienergy Brillouin zone [$-\pi/T$,$\pi/T$] \cite{PhysRevLett.113.236803}. Topological semimetals of periodically driven system are defined in such quasienergy spectrum \cite{PhysRevB.94.075443}. The core feature of these phases is the symmetry-protected band crossings. Inspired by the result in Ref. \cite{PhysRevB.93.184306,PhysRevB.110.235140}, we can further reach the general conditions for the appearance of band touching points. When $[\mathcal{H}(t),\mathcal{H}(t')]=0,\,\,\,\,{\forall}\in t,t'$, the one-period evolution operator is
\begin{equation}
U(T)=e^{-i\mathcal{H}_{eff}T}=e^{-i\int_0^{T} \mathcal{H}(t)dt}.
\end{equation}
The eigenvalue of $U(T)$ is $\mathcal{E} T=e^{-i\int_0^{T} E(t)dt}$, with $E(t)$ being eigenvalue of $\mathcal{H}(t)$. We can obtain that the quasienergy is $\mathcal{E}=\frac{1}{T}\int_0^{T}E(t)dt$. So band touching points occur when
\begin{eqnarray}
\begin{cases}
[\mathcal{H}(t),\mathcal{H}(t')]=0,\,\,\,\,{\forall}\in t,t'\\
\int_0^{T}E(t)\text{d}t=C\pi,~C\in\mathbb{Z}
\end{cases}\label{hh1}
\end{eqnarray}
at the quasienergy zero (or $\pi/T$) if $C$ is even (or odd). This provides a foundation for subsequent research of topological phase.

\section{Floquet composite Dirac semimetals} 
Here, we consider a periodically driven system whose bloch Hamiltonian is given by
\begin{equation}
\mathcal{H}(\mathbf{k},t)=\mathcal{H}_1(\mathbf{k})+\sum_n\mathcal{H}_2(\mathbf{k})\delta(t/T-n),
\end{equation}
where $\mathcal{H}_1({\bf k})=\sum_{i=x,y}[(\cos k_i+\lambda)^2-\sin^2k_i]s_0\sigma_z+[2(\cos k_x+\lambda)\sin k_y-2(\cos k_y+\lambda)\sin k_x  ]s_0\sigma_y+[\sum_{i=x,y} 2(\cos k_i+\lambda)\sin k_i]s_z\sigma_x$, $\mathcal{H}_2=(t_1\cos k_z+t_2)s_0\sigma_z$, $T$ is the driving period, and $n$ is an integer. $\sigma_i$ ($s_i$) are Pauli matrices acting on orbital (spin) degrees of freedom. $\mathcal{H}_1(\mathbf{k})$ and $\mathcal{H}_2(\mathbf{k})$ describe a second-order topological insulator and a semimetal, respectively. Both $\mathcal{H}_1({\bf k})$ and $\mathcal{H}_2({\bf k})$ possess inversion symmetry and time-reversal symmetry, with the corresponding symmetry operators given by $\mathcal{P} = s_0 \sigma_z$ and  $\mathcal{T} = i s_2 \sigma_0 \mathcal{K}$, respectively \cite{PhysRevLett.123.177001}. $\sigma_0$ and $s_2$ are Pauli matrices acting on degrees of freedom for orbital and spin, respectively. Here, $\mathcal{K}$ is complex conjugation operator. However, the effective Hamiltonian  $\mathcal{H}_{eff}=\frac{i}{T}\ln[e^{-i\mathcal{H}_2(\mathbf{k})T}e^{-i\mathcal{H}_1(\mathbf{k})T}]$ lacks the symmetries of the original static system. This makes it hard to study Dirac semimetal in our Floquet system. We propose
the following scheme to resolve this problem. All the symmetries can be inherited by $\mathcal{H}_{eff}$ in symmetric time frame. This frame is obtained by shifting the starting time of the evolution backward over half of the driving period. The resulting new effective Hamiltonian is $\mathcal{H}'_{eff}=\frac{i}{T}\ln[e^{-i\mathcal{H}_1(\mathbf{k})T/2}e^{-i\mathcal{H}_2(\mathbf{k})T}e^{-i\mathcal{H}_1(\mathbf{k})T/2}]$ \cite{PhysRevB.110.235140}. This Floquet system may provide a platform for studying Dirac semimetals.

Beyond the second-order topology found in the static case, periodic driving in our system gives rise to emergent first-order topology. To characterize the $k_z$-dependent first- and second-order topological phases, appropriate topological invariants need to be identified. Due to the time-reversal, space-inversion, and rotation chiral ($s_0\sigma_y\mathcal{H}'_{eff}(k_x,k_y,k_z)s_0\sigma_y=-\mathcal{H}'_{eff}(k_y,k_x,k_z)$) symmetries, the overall topology of the topological phases at the quasienergy $\alpha/T$, with $\alpha=0$ or $\pi$, is described by spin winding numbers $\mathcal{W}_{\alpha/T}(k_z)=[\mathcal{W}_1(k_z)+e^{i\alpha}\mathcal{W}_2(k_z)]/2$, where $\mathcal{W}_1(k_z)$ and $\mathcal{W}_2(k_z)$ are the mirror spin winding numbers defined in $\tilde{\mathcal{H}}_{1}({\bf k})=iT^{-1}\ln [e^{-i\mathcal{H}_1(\mathbf{k})T/2}e^{-i\mathcal{H}_2(\mathbf{k})T}e^{-i\mathcal{H}_1(\mathbf{k})T/2}]$ and $\tilde{\mathcal{H}}_{2}({\bf k})=iT^{-1}\ln [e^{-i\mathcal{H}_2(\mathbf{k})T/2}e^{-i\mathcal{H}_1(\mathbf{k})T}$$e^{-i\mathcal{H}_2(\mathbf{k})T/2}]$ with symmetric time frame, along the high-symmetry line $k_x=k_y=k$ \cite{PhysRevB.90.125143,PhysRevLett.123.177001}. 

The first-order topology of static systems with the same symmetries can be described by the spin Chern number \cite{PhysRevB.108.075138}. For our periodically driven system, the spin Chern number can be extended to dynamical spin winding number \cite{PhysRevB.104.L180303}. This can be calculated via the following approach. First, $\mathcal{H}(\mathbf{k},t)$ can be rewritten as Diag[$\mathcal{H}_{+}(\mathbf{k},t),\mathcal{H}_{-}(\mathbf{k},t)$]. Here, $\mathcal{H}_\pm({\bf k},t)=\sum_{i=x,y}[(\cos k_i+\lambda)^2-\sin^2k_i]\sigma_z+[2(\cos k_x+\lambda)\sin k_y-2(\cos k_y+\lambda)\sin k_x  ]\sigma_y\pm[\sum_{i=x,y}2(\cos k_i+\lambda)\sin k_i]\sigma_x+\sum_n(t_1\cos k_z+t_2)\sigma_z\delta(t/T-n)$. Then the dynamical spin winding number is defined as $\mathcal{V}_{\alpha/T}(k_z)=[\mathcal{V}_{\alpha/T,+}(k_z)-\mathcal{V}_{\alpha/T,-}(k_z)]/2$ with
\begin{eqnarray}
\mathcal{V}_{\alpha/T,\pm}&(k_z)&=\frac{1}{24\pi^2}\int_{0}^{T}dt\int_0^{2\pi}dk_x\int_0^{2\pi} dk_y\epsilon^{i_1i_2i_3}\nonumber\\&&{\rm tr}[u^{\dag}_{\alpha,\pm}\partial_{i_1} u_{\alpha,\pm} u^{\dag}_{\alpha,\pm}\partial_{i_2} u_{\alpha,\pm} u^{\dag}_{\alpha,\pm}\partial_{i_3} u_{\alpha,\pm}]\textbf{}.
\end{eqnarray}
Here $\epsilon_{ijk}$, with $\{i,j,k\}\in \{t, k_x,k_y\}$, is the completely antisymmetric tensor and a sum over repeated indices has been used, ${u}_{\alpha,\pm}=\tilde{U}_{\pm}(t)\sum_{l=1}^2e^{i\mathcal{E}^{(\alpha)}_{l,{\bf k}}t}$$|u_{l,\pm}({\bf k},T)\rangle\langle u_{l,\pm}({\bf k},T)|$, where the $\tilde{U}_{\pm}(t)=\mathbb{T}e^{-i\int_{0}^{t}\mathcal{H}_{\pm}(\mathbf{k},t)dt}$ and $\mathcal{E}^{(\alpha)}_{l,\mathbf{k}}$ is the quasienergy in the $l$th band chosen in the regime $[\alpha/T,(\alpha +2\pi)/T)$, $\mathcal{E}^{(\alpha)}_{l,\mathbf{k}}$ and $|u_{l,\pm}({\bf k},T)\rangle$ is the eigenvalue and corresponding eigenstate of $\frac{i}{T}\ln[\tilde{U}_{\pm}(T)]$. $\mathcal{V}_{\alpha/T,\pm}(k_z)$ can be used to count the number of gapless edge states.  $\mathcal{V}_{\alpha/T}(k_z)$ is the $Z_2$ topological invariant \cite{PhysRevB.96.195303}. If $\mathcal{V}_{\alpha/T}(k_z)$ is an odd number, there is first-order gapless edge states at $\alpha/T$.

{Inspired by the study of the static model with second-order topology in Ref. \cite{PhysRevLett.123.177001},  we construct a similar description in our Floquet system. For the static model in Ref. \cite{PhysRevLett.123.177001}, spin winding numbers $\mathcal{W}$ defined on the high-symmetry line $k_x=k_y$ (or $k_x=-k_y$) can be used as a topological number. 
 When the system with non-trivial winding number takes a ribbon geometry with open boundary condition in the $\hat{x}+\hat{y}$ (or $\hat{x}-\hat{y}$) direction, gapless modes appear on this edge. However, due to that the Chern number of bulk band is zero, the number of left-moving modes and right-moving modes must be equal. These gapless edge modes are expected to be gapped when the open boundary condition is along other direction. The Dirac mass gapping out the gapless edge modes will
have opposite sign if the respective edges are located at
different sides of the $\hat{x}+\hat{y}$ (or $\hat{x}-\hat{y}$) direction \cite{PhysRevLett.123.177001}. Therefore, 0-mode corner states can appear at the intersections of these edges. The number of corner states is $4\mathcal{W}$. This mechanism remains valid in our Floquet system. To describe our periodically driven system, the topological invariant that characterizes second-order topology can be generalized as $|\mathcal{W}_{\alpha/T}(k_z)|-|\mathcal{V}_{\alpha/T}(k_z)|$. The number of the second-order corner states at the quasienergy $\alpha/T$ equals to $4(|\mathcal{W}_{\alpha/T}(k_z)|-|\mathcal{V}_{\alpha/T}(k_z)|)$ \cite{PhysRevLett.123.177001,PhysRevB.107.235132}.}

To illustrate composite Dirac semimetals, we consider an example with parameters $\lambda=0.3$, $t_1=1.5$, $t_2=1.6$, and $T=1$. The defining characteristic of Dirac semimetals is the presence of band-touching points, known as Dirac points. From Eq. \eqref{hh1}, we derive the band-touching condition:
\\ \textbf{Case I:} $k_x=k_y=0$, then Eq. \eqref{hh1} requires
\begin{equation}
2T(1+\lambda)^2+T(t_1\cos k_z+t_2)=C\pi, \,\,C=2. \label{smn1}
\end{equation}
we can find that there are two Dirac points ($k_x$, $k_y$, $k_z$)=(0, 0, $\pm$ 0.518) in the zero gap.\\
\textbf{Case II:} $k_x=k_y=\pi$, then Eq. \eqref{hh1} requires
\begin{equation}
2T(\lambda-1)^2+T(t_1\cos k_z+t_2)=C\pi, \,\,C=1. \label{smn2}
\end{equation}
we can find that there are two Dirac points ($k_x$, $k_y$, $k_z$)=($\pi$, $\pi$, $\pm$ 1.1871) in the $\pi/T$ gap.\\
\textbf{Case III:} $k_x=0$, $k_y=\pi$ ($k_x=\pi$, $k_y=0$), then Eq. \eqref{hh1} requires
\begin{equation}
T[(\lambda+1)^2+(\lambda-1)^2+(t_1\cos k_z+t_2)]=C\pi, \,\,C=1. \label{smn3}
\end{equation}
we can find that there are four Dirac points ($k_x$, $k_y$, $k_z$)=($0$, $\pi$, $\pm$ 2.0104) and ($\pi$, $0$, $\pm$ 2.0104) in the $\pi/T$ gap.\\
\textbf{Case IV:} $k_x=\arccos{(-\lambda)}$, $k_y=-\arccos{(-\lambda)}$, then Eq. \eqref{hh1} requires
\begin{equation}
-2T(1-\lambda^2)+T(t_1\cos k_z+t_2)=C\pi, \,\,C=0. \label{smn4}
\end{equation}
we can find that there are eight Dirac points ($k_x$, $k_y$, $k_z$)=($\pm\arccos{(-\lambda)}$, $\pm\arccos{(-\lambda)}$, $\pm$ 1.4236) in the zero gap. Numerical results show that these band-touching points all exhibit linear dispersion relations in their vicinity, thus identifying them as Dirac points. Due to the rich results, both the number and location of Dirac points are tunable.

\begin{figure}[tbp]
\centering
\includegraphics[width=1\columnwidth]{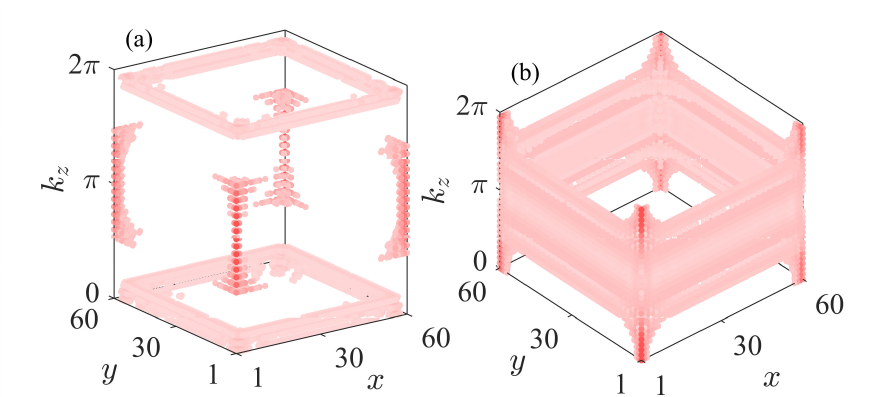}
\caption{The probability distributions of (a) zero- and (b) $\pi/T$-mode topological states in different $k_z$-dependent 2D subsystem. We use $\lambda=0.3$, $t_1=1.5$, $t_2=1.6$, and $T=1$.} \label{traj2}
\end{figure}

\begin{figure}[tbp]
\centering
\includegraphics[width=1\columnwidth]{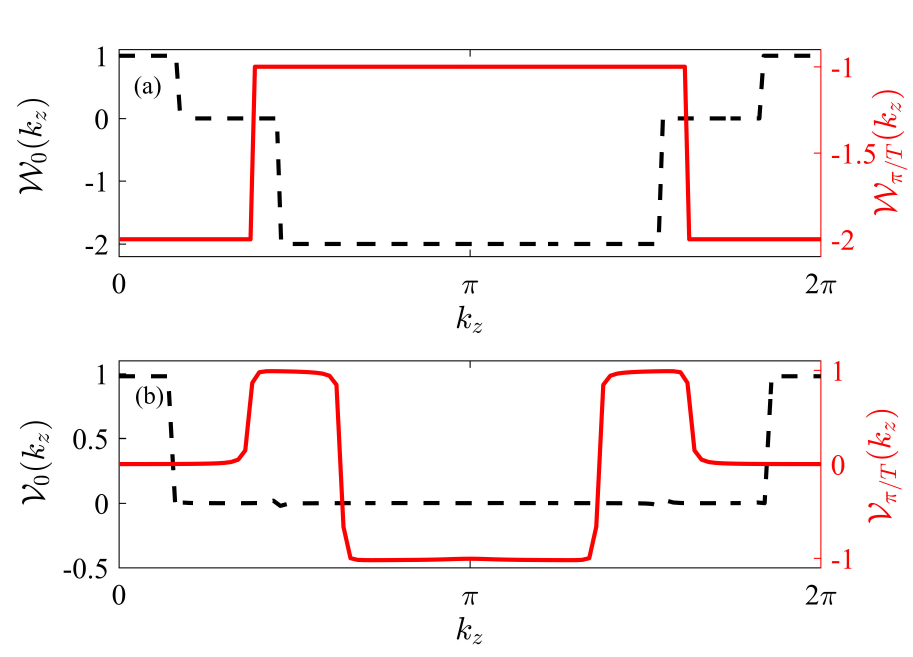}
\caption{(a) $\mathcal{W}_{\alpha/T}$ and (b) $\mathcal{V}_{\alpha/T}$ with the change of $k_z$. We use $\lambda=0.3$, $t_1=1.5$, $t_2=1.6$, and $T=1$. } \label{trajtv}
\end{figure}

Fig. \ref{traj2} displays the probability distributions of boundary states. We can observe three typical regimes from  zero gap: 
\\  
\text{(i)} An open zero-mode gap for $0.518 < k_z < 1.4236$ and $4.8596 < k_z < 5.7652$ . \\  
\text{(ii)} Gapless edge states for $0 \leq k_z < 0.518$ and $5.7652 < k_z \leq 2\pi$.\\  
\text{(iii)} Zero-mode corner states for $1.4236 < k_z < 4.8596$ . \\  
As indicated by the topological invariants in Fig. \ref{trajtv}, the aforementioned three regimes correspond to: (i) trivial, (ii) first-order, and (iii) second-order $k_z$-dependent insulators. Phase transitions between trivial and first/second-order topological insulators in $k_z$-dependent 2D subsystems, occurring at gap-closing points, manifest as type-I (type-II) Dirac points. The edge (corner) states of first (second)-order topological insulators contribute to surface (hinge) Fermi arcs that connect a pair of Dirac points. Similarly, there are type III Dirac points at $\pi/T$ gap. This type-III Dirac point gives rise to coexisting surface and hinge Fermi arcs.  However, it is noted that the Dirac points at $k_z=\pm2.0104$ separate the same first-order topological insulator. Here, the location of the Dirac points agrees with the conclusions from our analytical approach presented earlier. This confirms that our periodic driving protocol generates composite Dirac semimetals featuring coexisting type-I, type-II, and type-III Dirac points. Although the Ref. \cite{PhysRevB.100.161401} reported the composite Dirac semimetals,  it's a
 stable combination of a weak  topological insulator and a Dirac semimetal. Such results in our work have not been discovered yet. 

\begin{figure}[tbp]
\centering
\includegraphics[width=1\columnwidth]{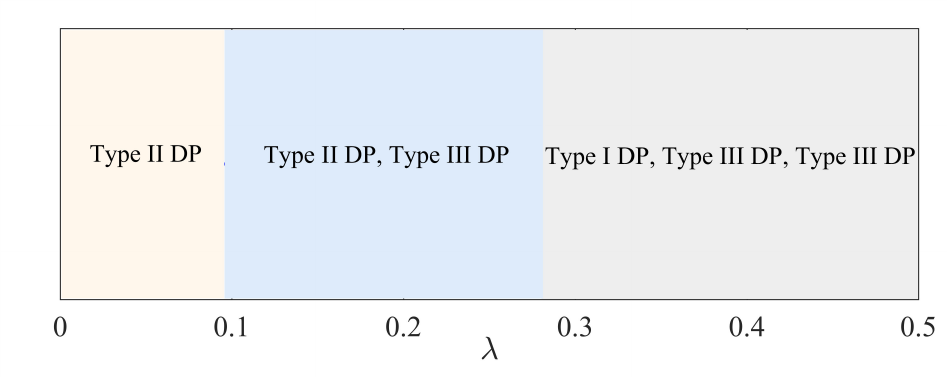}
\caption{Phase diagram. We use $t_1=0.75$, $t_2=2.25$, and $T=1$.} \label{traj3} 
\end{figure}

In order to give a global picture of Dirac semimetals in our Floquet system, we plot in Fig. \ref{traj3} the phase diagram. We identified three distinct Dirac semimetal phases: (i) a phase with type II Dirac points, (ii) a composite phase with coexisting type II and type III Dirac points, and (iii) a phase hosting all three types of Dirac points (I, II, and III). Each phase transition in Fig. \ref{traj3} is accompanied by the generation of a pair of new Dirac points.  Combining with the previous Eq. \eqref{hh1}, we can identify the phase boundaries at $\lambda = 0.0940$ and $\lambda = 0.2812$. The numerical results in Fig. \ref{traj3} perfectly obey our theory. This result indicates that periodic driving is a useful tool for exploring new topological phases absent in its static counterpart. 

 It is noted that the delta-function driving protocol is considered for simplicity. Our scheme is generalizable to other driving forms, such as harmnonic driving
\begin{equation}
\mathcal{H}(\mathbf{k},t)=\mathcal{H}_1(\mathbf{k})+t_1[\cos k_z+\cos (wt)]s_0\sigma_z,
\end{equation}
where $w$ is the frequency of periodic driving. Owing to the coexistence of inversion and time-reversal symmetries, this Floquet system provides a platform for studying Dirac semimetals.

Here, we choose $\lambda=0.3$, $t_1=3$, and $\omega=2\pi$. The Dirac points can also be obtained by Eq. \eqref{hh1}.
\\ \textbf{Case I:} $k_x=k_y=0$, then Eq. \eqref{hh1} requires
\begin{equation}
2T(1+\lambda)^2+Tt_1\cos k_z=C\pi, \,\,C=1,2. \label{smn5}
\end{equation}
we can find that there are two Dirac points ($k_x$, $k_y$, $k_z$)=(0, 0, $\pm$ 0.2547) in the zero gap and two Dirac points ($k_x$, $k_y$, $k_z$)=(0, 0, $\pm$ 1.6503) in the $\pi/T$ gap.\\
\textbf{Case II:} $k_x=k_y=\pi$, then Eq. \eqref{hh1} requires
\begin{equation}
2T(\lambda-1)^2+Tt_1\cos k_z=C\pi, \,\,C=0,1. \label{smn6}
\end{equation}
we can find that there are two Dirac points ($k_x$, $k_y$, $k_z$)=($\pi$, $\pi$, $\pm$ 1.9036) in the zero gap and two Dirac points ($k_x$, $k_y$, $k_z$)=($\pi$, $\pi$, $\pm$ 0.7662) in the $\pi/T$ gap.\\
\textbf{Case III:} $k_x=0$, $k_y=\pi$ ($k_x=\pi$, $k_y=0$), then Eq. \eqref{hh1} requires
\begin{equation}
T(\lambda+1)^2+T(\lambda-1)^2+Tt_1\cos k_z=C\pi, \,\,C=0,1. \label{smn7}
\end{equation}
we can find that there are four Dirac points ($k_x$, $k_y$, $k_z$)=($0$, $\pi$, $\pm$ 2.3843) and ($\pi$, $0$, $\pm$ 2.3843) in the zero gap and four Dirac points ($k_x$, $k_y$, $k_z$)=($0$, $\pi$, $\pm$ 1.2445) and ($\pi$, $0$, $\pm$ 1.2445) in the $\pi/T$ gap.\\
\textbf{Case IV:} $k_x=\arccos{(-\lambda)}$, $k_y=-\arccos{(-\lambda)}$, then Eq. \eqref{hh1} requires
\begin{equation}
-2T(1-\lambda^2)+Tt_1\cos k_z=C\pi, \,\,C=0,-1. \label{smn8}
\end{equation}
we can find that there are eight Dirac points ($k_x$, $k_y$, $k_z$)=($\pm\arccos{(-\lambda)}$, $\pm\arccos{(-\lambda)}$, $\pm$ 0.9189) in the zero gap and eight Dirac points ($k_x$, $k_y$, $k_z$)=($\pm\arccos{(-\lambda)}$, $\pm\arccos{(-\lambda)}$, $\pm$ 2.0270) in the $\pi/T$ gap. There are thirty two Dirac points in Brillouin zone. The periodic driving enriches the controllability of Dirac points.


\begin{figure}[tbp]
\centering
\includegraphics[width=1\columnwidth]{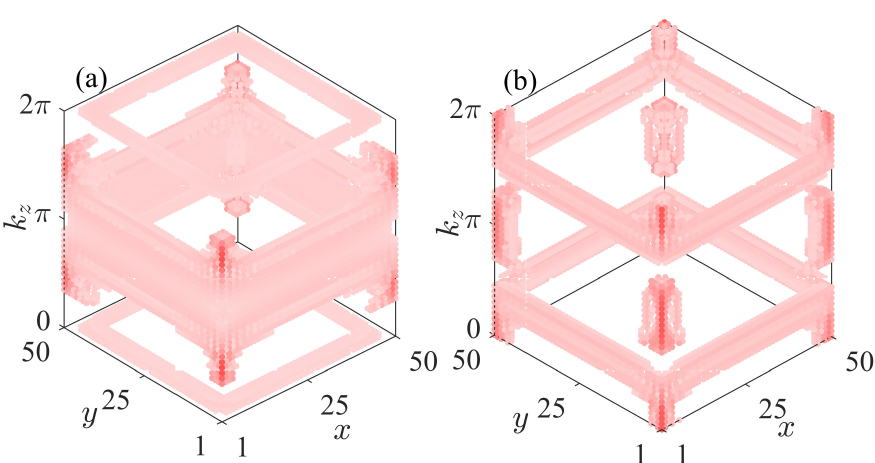}
\caption{The probability distributions of (a) zero- and (b) $\pi/T$-mode topological states in different $k_z$-dependent 2D subsystem. We use $\lambda=0.3$, $t_1=3$, and $w=2\pi$.  } \label{traj4}
\end{figure}

Fig. \ref{traj4} displays the probability distributions of zero- and $\pi/T$-mode topological states in $k_z$-dependent subsystems. The critical points that separate the different $k_z$-dependent topological insulators are just analytical Dirac points. We observe that there are type I, II, and III Dirac points in both zero and $\pi/T$ gaps. This represents a more comprehensive result. Correspondingly, there are coexisting surface and hinge Fermi arcs in both gaps. Therefore, our proposal suggests a degree of universality. These enable novel applications of composite Dirac semimetals.

\section{Discussion and conclusion} In recent years, Dirac semimetals have been observed in sonic crystals \cite{PhysRevLett.127.146601}, Cd$_{3}$As$_{2}$ \cite{PhysRevLett.113.027603,PhysRevLett.120.016801}, Au$_2$Pb \cite{PhysRevLett.130.236402}, and nitrogen-vacancy center in diamond \cite{PhysRevLett.134.153601}. Meanwhile, periodic driving has exhibited its {versatility} in engineering exotic phases {across} various experimental platforms, including ultracold atoms \cite{RevModPhys.89.011004,PhysRevLett.116.205301,PhysRevLett.130.043201}, thermal Atoms \cite{PhysRevLett.133.133601}, superconductor qubits \cite{Roushan2017,PhysRevLett.134.090402}, quantum
 materials \cite{PhysRevLett.131.116401}, photonics \cite{Rechtsman2013,PhysRevLett.122.173901,pan2023realhigherorderweylphotonic,PhysRevLett.133.073803}, acoustic system \cite{PhysRevLett.129.254301,PhysRevLett.134.126603}. Based on these developments, we believe that our approach is experimentally feasible.

 In summary, we have demonstrated periodic-driving induced composite Dirac semimetals in a four-band system with time-reversal and space-inversion symmetries. A general description is established to characterize Dirac semimetals in Floquet systems. Both the location of Dirac points and the topological invariant can be determined within the theoretical framework. The numerical results indicate that composite Dirac semimetals with the coexistence of type I, II, and III Dirac points can be induced by delta-function or harmonic driving. Such result has not been reported before. This significantly expands the scope of the topological materials and enriches their controllability.

\section{Acknowledgments}We would like to acknowledge helpful discussions with Ken Chen. This work is supported by National Natural Science Foundation (Grants No. 12405007 and NO. 12305011), Funds for Young Scientists of Chongqing Municipal Education Commission(Grant No. KJQN20240 and N0. KJQN202500619), Natural Science Foundation of Chongqing (Grant No. CSTB2025NSCQ-GPX1265, No. CSTB2022NSCQ-MSX0316, No. CSTB2024NSCQ-MSX0736, No. CSTB2025NSCQ-GPX1272), startup grant at CQUPT (Grant 
 No. E012A2020016, E012A2022017 and E012A2024044), and the research foundation of the Institute for Advanced Sciences of CQUPT (Grant No. E011A2022328).

\bibliography{references}
\end{document}